\documentstyle[preprint,tighten,prl,aps]{revtex}

\begin{document}
\preprint{IFUP-TH 36/94}
\title{Asymptotic scaling from strong coupling}
\author{Massimo Campostrini, Paolo Rossi, and Ettore Vicari}
\address{Dipartimento di Fisica dell'Universit\`a and I.N.F.N.,
I-56126 Pisa, Italy}
\maketitle
\begin{abstract}
Strong-coupling analysis of two-dimensional chiral models, extended to
15th order, allows for the identification of a scaling region where
known continuum results are reproduced with great accuracy, and
asymptotic scaling predictions are fulfilled.
The properties of the large-$N$ second-order phase transition are
quantitatively investigated.
\end{abstract}
\pacs{11.15.Me, 11.15.Pg}

\narrowtext

Recent developments in the analytical and numerical study of lattice
two-dimensional ${\rm SU}(N)\times{\rm SU}(N)$ principal chiral models
have shown the existence of a scaling region, where continuum
predictions for dimensionless ratios of physical quantities are
substantially verified \cite{chiral}.

The scaling region begins at very small values of the correlation
length: $\xi\gtrsim2$, well within the region of convergence of the
strong-coupling series.  Moreover by performing a nonperturbative
change of variables \cite{Parisi} from the standard Lagrangian
(inverse) coupling $\beta$ to
\begin{equation}
\beta_E = {N^2 - 1 \over 8 N^2 E},
\label{betaE}
\end{equation}
where $E$ is the internal energy of the system (which can be measured
in a Monte Carlo simulation), one can find agreement in the whole
scaling region between the measured mass scale and the prediction for
the mass gap -- $\Lambda$ parameter ratio offered by the two-loop
continuum renormalization group supplemented by a Bethe-Ansatz
evaluation \cite{Balog}.  As a matter of fact, this may be thought as
evidence for asymptotic scaling within the strong-coupling regime.

We therefore felt motivated for a significant effort in improving our
analytical knowledge of the strong-coupling series for principal
chiral models, in order to check the accuracy of predictions for
physical (continuum) quantities solely based on strong-coupling
computations.

As a byproduct, strong-coupling series can be analyzed to investigate
the critical behavior of the $N=\infty$ theory, where Monte Carlo data
seem to indicate the existence of a transition at finite $\beta$.

Strong-coupling calculations for matrix models are best performed by
means of the character expansion.
Even the character expansion however eventually runs into almost
intractable technical difficulties, related to two concurrent
phenomena:

\begin{enumerate}
\item  the proliferation of configurations contributing to large orders
    of the series, whose number essentially grows like that of
    non-backtracking random walks;
\item  the appearance of topologically nontrivial configurations
    corresponding to group integrations that cannot be performed by
    applying the orthogonality of characters and the decomposition of
    their products into sums.
\end{enumerate}

We shall devote an extended paper to a complete presentation of our
15th-order strong-coupling character expansions for arbitrary
${\rm U}(N)$ groups (extension to ${\rm SU}(N)$ is then obtained by
the techniques discussed in Refs.~\cite{Green} and \cite{chiral}).  We
only mention that we had to rely on a mixed approach: problem 1 above
was tackled by a computer program generating all non-backtracking
random walks, computing their correct multiplicities, and classifying
them according to their topology; problem 2, requiring the analytical
evaluation of many classes of nontrivial group integrals, and
extraction of their connected contribution to the relevant physical
quantities, was solved by more conventional algebraic techniques.

We pushed our computational techniques close to their limit; in order
to extend our results to higher orders, a more algorithmic approach
would be in order, especially for recognition of diagram topologies
and group integration.

In the present letter we shall only exhibit the results we obtained by
taking the large-$N$ limit of all the quantities we computed.
In the strong-coupling domain, convergence to $N=\infty$ is usually
fast, and therefore large-$N$ results are a good illustration of a
phenomenology that repeats itself (with small corrections) for finite
values of $N$.

Starting from the standard nearest-neighbour interaction
\begin{equation}
S_L = -2 N \beta \sum_{x,\mu}
{\mathop{\rm Re}}{\mathop{\rm Tr}}\{U(x)\,U^\dagger(x{+}\mu)\},
\end{equation}
we computed the large-$N$ free energy
\begin{equation}
F = {1\over N^2}\,\ln \int dU_n\exp(-S_L)
\label{F}
\end{equation}
to 18th order in the expansion in powers of $\beta$.  The resulting
series is
\begin{eqnarray}
F &=& \beta^2 + 2\beta^4 + 4\beta^6 + 19\beta^8 +
96\beta^{10} + 604\beta^{12} + 4036\beta^{14}
\nonumber \\ &&\quad+\;
{58471 \over 2}\beta^{16} + {663568 \over 3}\beta^{18} +
{\rm O}\bigl(\beta^{20}\bigr).
\end{eqnarray}

The two-point Green's functions, normalized to obtain a finite
large-$N$ limit, are
\begin{equation}
G(x) = {1\over N}\,{\mathop{\rm Tr}}\{U^\dagger(x)U(0)\}.
\end{equation}
We computed all nontrivial two-point Green's functions up to
${\rm O}(\beta^{15})$.  The corresponding information is appropriately
summarized by introducing the lattice Fourier transform
\begin{equation}
\tilde G(p) = \sum_x G(x) \exp(i p\cdot x)
\end{equation}
and evaluating the inverse propagator
\widetext
\begin{eqnarray}
{\tilde G}^{-1}(p) &=& A_0 + {{\hat p}^2} A_1 +
\bigl({{\bigl({{\hat p}^2}\bigr)}^2} - {{\hat p}^4}\bigr) A_{2,0} +
{{\hat p}^4} A_{2,2} +
{{\hat p}^2}\bigl({{\bigl({{\hat p}^2}\bigr)}^2} -
        {{\hat p}^4}\bigr) A_{3,1} +
{{\hat p}^6} A_{3,3}
\nonumber \\ &&\quad+\;
%
{\bigl({{\bigl({{\hat p}^2}\bigr)}^2} - {{\hat p}^4}\bigr)}^2 A_{4,0}
+
%
{{\hat p}^4} \bigl({{\bigl({{\hat p}^2}\bigr)}^2} -
        {{\hat p}^4}\bigr) A_{4,2} +
{{\hat p}^8} A_{4,4} +
{{\hat p}^2}{{\bigl({{\bigl({{\hat p}^2}\bigr)}^2} -
        {{\hat p}^4}\bigr)}^2} A_{5,1}
\nonumber \\ &&\quad+\;
%
{{\hat p}^6} \bigl({{\bigl({{\hat p}^2}\bigr)}^2} -
        {{\hat p}^4}\bigr) A_{5,3} + ... \, ,
\end{eqnarray}
where
\begin{eqnarray}
A_0 &=& 1 - 4\beta + 4\beta^2 - 4\beta^3 + 12\beta^4 -
28\beta^5 + 52\beta^6 - 132\beta^7 + 324\beta^8 -
908\beta^9 + 2020\beta^{10}
\nonumber \\ &&\quad-\;
6284\beta^{11} + 15284\beta^{12} - 48940\beta^{13} +
116596\beta^{14} - 393052\beta^{15} +
{\rm O}\bigl(\beta^{16}\bigr) ,
\nonumber \\
A_1 &=& \beta\bigl( 1 + \beta^2 + 7\beta^4 + 4\beta^5 +
33\beta^6 + 32\beta^7 + 243\beta^8 + 324\beta^9 +
1819\beta^{10} + 2520\beta^{11}
\nonumber \\ &&\quad+\;
14859\beta^{12} + 23124\beta^{13} + 123767\beta^{14} \bigr) +
{\rm O}\bigl(\beta^{16}\bigr) ,
\nonumber \\
A_{2,0} &=& -\beta^6\left( 1 + 6\beta^2 + 8\beta^3 + 57\beta^4
+ 116\beta^5 + 500\beta^6 + 1152\beta^7 + 5173\beta^8 +
11600\beta^9 \right)
\nonumber \\ &&\quad+\;
%
 {\rm O}\bigl(\beta^{16}\bigr) ,
\nonumber \\
A_{2,2} &=& -2\beta^8\left( 1 + 2\beta + 12\beta^2 + 35\beta^3 +
121\beta^4 + 408\beta^5 + 1424\beta^6 + 4244\beta^7 \right) +
{\rm O}\bigl(\beta^{16}\bigr) ,
\nonumber \\
A_{3,1} &=& \beta^9\left(1 + {29 \over 2} \beta^2 + 26\beta^3 +
144\beta^4 + 488\beta^5 + 1802\beta^6 \right) +
{\rm O}\bigl(\beta^{16}\bigr) ,
\nonumber \\
A_{3,3} &=& 2\beta^{11}\left( 1 + 2\beta + 20\beta^2 + 72\beta^3
+ 272\beta^4 \right) + {\rm O}\bigl(\beta^{16}\bigr) ,
\nonumber \\
A_{4,0} &=& - \beta^{12}\left({5 \over 2} + 37\beta^2 +
84\beta^3 \right) + {\rm O}\bigl(\beta^{16}\bigr) ,
\nonumber \\
A_{4,2} &=& - \beta^{12}\left( 1 + 32\beta^2 + 64\beta^3 \right)
+ {\rm O}\bigl(\beta^{16}\bigr) ,
\nonumber \\
A_{4,4} &=& -2\beta^{14}\left( 1 + 2\beta \right) +
{\rm O}\bigl(\beta^{16}\bigr) ,
\nonumber \\
A_{5,1} &=& 7\beta^{15} + {\rm O}\bigl(\beta^{16}\bigr) ,
\nonumber \\
A_{5,3} &=& \beta^{15} + {\rm O}\bigl(\beta^{16}\bigr) .
\end{eqnarray}

A number of physically interesting quantities can be extracted from
${\tilde G}^{-1}(p)$.  In particular the magnetic susceptibility is
\begin{eqnarray}
\chi = \sum_x G(x) = {1\over A_0} &=&
1 + 4\beta + 12\beta^2 + 36\beta^3 + 100\beta^4 +
284\beta^5 + 796\beta^6 + 2276\beta^7 + 6444\beta^8
\nonumber \\ &&\quad+\;
%
 18572\beta^9 + 53292\beta^{10}
+
%
155500\beta^{11} + 451516\beta^{12} + 1330796\beta^{13}
\nonumber \\ &&\quad+\;
%
3904924\beta^{14} + 11617404\beta^{15}
+ {\rm O}\bigl(\beta^{16}\bigr) ,
\end{eqnarray}
while the second-moment definition of the correlation length leads to
\begin{eqnarray}
M^2_G = {1\over\left<x^2\right>_G} = {A_0\over A_1} &=&
{1\over \beta} - 4 + 3\,\beta + 2\,\beta^3 - 4\,\beta^4 +
12\,\beta^5 - 40\,\beta^6 + 84\,\beta^7 - 296\,\beta^8 +
550\,\beta^9
\nonumber \\ &&\quad-\;
1904\,\beta^{10} + 3316\,\beta^{11} - 15288\,\beta^{12} +
28016\,\beta^{13} + {\rm O}\bigl(\beta^{14}\bigr) .
\end{eqnarray}
and the corresponding wavefunction renormalization is
$Z_G = A_1^{-1}$.

Moreover, by solving the equation
\begin{equation}
{\tilde G}^{-1}(p_x{=}i\mu_{\rm s}, p_y{=}0) = 0,
\label{Gs}
\end{equation}
we can compute the wall-wall correlation length, i.e.\ the true
mass-gap $\mu_{\rm s}$; Eq.~(\ref{Gs}) is algebraic in
\begin{eqnarray}
M^2_{\rm s} = 2(\cosh\mu_{\rm s} - 1) &=&
{1\over\beta} - 4 + 3\beta + 2\beta^3 - 4\beta^4 + 10\beta^5
-
28\beta^6 + 48\beta^7 - 206\beta^8
\nonumber \\ &&\quad+\;
%
352\beta^9 - 1510\beta^{10} + 2354\beta^{11}
+ {\rm O}\bigl(\beta^{12}\bigr) .
\end{eqnarray}
\narrowtext
By solving
\begin{equation}
{\tilde G}^{-1}(p_x{=}i\mu_{\rm d}/\sqrt{2}, p_y{=}i\mu_{\rm d}/\sqrt{2}) = 0
\end{equation}
we find the diagonal wall-wall correlation length
\begin{eqnarray}
M^2_{\rm d} &=& 4\left(\cosh{\mu_{\rm d}\over\sqrt2} - 1\right) =
{1\over \beta} - 4  + 3\beta + {3\over 2}\beta^3 - 3\beta^5
-
%
{{169\over 8}\beta^7} - {{881}\over 4}\beta^9
+ {\rm O}\bigl(\beta^{11}\bigr) .
\end{eqnarray}
In order to obtain the highest-order contribution to $M^2_{\rm s}$, it
was necessary to compute a few (long-distance) Green's functions to
16th and 17th order.

Lattice chiral models have a peak in the specific heat
\begin{equation}
C={1\over N}\,{dE\over dT}\,,\qquad
T = {1\over N\beta}\,,
\end{equation}

which becomes more and more pronounced with increasing $N$
\cite{chiral}.  Fig.~\ref{Cdata} shows Monte Carlo data for the
specific heat of ${\rm SU}(N)$ models for $N=21,30$, and ${\rm U}(N)$
models for $N=15,21$.  (We recall that ${\rm U}(N)$ models at finite
$N$ should experience a phase transition with a pattern similar to the
$\rm XY$ model, but its location is beyond the specific heat peak).
With increasing $N$, the positions of the peaks in ${\rm SU}(N)$ and
${\rm U}(N)$ seems to approach the same value of $\beta$, consistently
with the fact that ${\rm SU}(N)$ and ${\rm U}(N)$ models must have the
same large-$N$ limit.  This should be considered an indication of a
phase transition at $N=\infty$; a rough extrapolation of the $C$ data
indicates a critical coupling $\beta_c\simeq 0.306$.
Extrapolating the values of $\xi_G=1/M_G$ and $\beta_E$ at the peak of
$C$ to $N=\infty$, we obtain respectively $\xi_G^{(c)} \simeq 2.8$
and $\beta_E^{(c)} \simeq 0.220$.

The above picture is confirmed by an analysis of the large-$N$ 18th
order strong-coupling series of $C$, based on the method of the
integral approximants \cite{Hunter,Fisher}.  We indeed obtained quite
stable results showing the critical behavior
\begin{equation}
C \sim  |\beta - \beta_c |^{-\alpha} \,,
\label{Ccrit}
\end{equation}
with $\beta_c \cong 0.3054$ and $\alpha \cong 0.21$, in agreement with
the extrapolation of Monte Carlo data.  Fig.~\ref{Cdata} shows that
simulation data of $C$ approach, for growing $N$, the strong-coupling
determination.

In spite of the existence of a phase transition at $N=\infty$, Monte
Carlo data show scaling and asymptotic scaling (in the $\beta_E$
scheme) even for $\beta$ smaller then the peak of the specific heat,
suggesting an effective decoupling of the modes responsible for the
phase transition from those determining the physical continuum limit;
this phenomenon motivated us to use the strong-coupling approach to
test scaling and asymptotic scaling.  In Fig.~\ref{scaling} we plot
the dimensionless ratio $\mu_s/M_G$ vs.\ the correlation length
$\xi_G\equiv 1/M_G$, as obtained from our strong-coupling series.
Notice the stability of the curve for a large region of values of
$\xi_G$ and the good agreement (within a few per mille) with the
continuum large-$N$ value extrapolated by Monte Carlo data
$\mu_s/M_G=0.991(1)$ \cite{chiral}.

In order to test asymptotic scaling we perform the variable change
indicated in Eq.(\ref{betaE}), evaluating the energy from its
strong-coupling series
\begin{equation}
E = 1 - {1 \over 4}\,{\partial F\over\partial\beta} = 1 - G((1,0)).
\label{E}
\end{equation}
(cf.\ Eq.~(\ref{F})).
The asymptotic scaling formula for the mass gap in the $\beta_E$
scheme is, in the large-$N$ limit,
\begin{eqnarray}
\mu &\cong& \sqrt{\pi\over e}\,16\exp\left(\pi\over4\right)
\Lambda_{E,2l}(\beta_E),
\qquad \Lambda_{E,2l}(\beta_E) =
%
\sqrt{8\pi\beta_E}\exp(-8\pi\beta_E);
\label{mass-lambda}
\end{eqnarray}
$\beta_E$ can be expressed as a strong-coupling expanded function
of $\beta$ by means of Eq.~(\ref{E}).
In Fig.~\ref{asySC} the strong-coupling estimates of
$\mu_s/\Lambda_{E,2l}$ and $M_G/\Lambda_{E,2l}$ are plotted vs.\
$\beta_E$, for a region of coupling corresponding to correlation
lengths $1\lesssim \xi_G \lesssim 3$.  These quantities agree with the
exact continuum prediction within 5\% in the whole region.

Since the large-$N$ $\beta$-function in the $\beta_E$ scheme is not
singular (as shown by Monte Carlo data \cite{chiral} and
strong-coupling analysis) and the specific heat has a divergence at
$\beta_c$, then the relationship
\begin{equation}
\beta_E(T_E) = {8N^2\over N^2-1}\, C(T) \beta_L(T)
\label{betafunction}
\end{equation}
between the $\beta$-function in the standard scheme $\beta_L(T)$ and
in the $\beta_E$ scheme $\beta_E(T_E)$ leads to an non-analytical
zero of $\beta_L(T)$ at $\beta_c$:
$\beta_L(T)\sim |\beta-\beta_c|^\alpha$, explaining the observed
behavior in $\beta$ of the Monte Carlo data for the mass gap at large
$N$ \cite{chiral}.

This phenomenon is further confirmed by an analysis of the
strong-coupling series of $\chi$ and $M^2_G$.  Assuming they are
well-behaved functions of the energy, we should have
\begin{equation}
{d \ln \chi\over d\beta} \sim  {d \ln M^2_G\over d\beta} \sim
|\beta-\beta_c|^{-\alpha}.
\label{chi_crit}
\end{equation}
Analyzing the corresponding series by a modified integral
approximant scheme which forces the approximant to have a singularity
at $\beta\simeq 0.3054$, we found a critical behavior consistent
with Eq.~(\ref{chi_crit}) ($\alpha\simeq0.2$).



\begin{figure}
\caption{Specific heat vs.\ $\beta$.  The solid line represents the
resummation of the strong-coupling series, whose estimate of the
critical $\beta$ is indicated by the vertical dashed lines.}
\label{Cdata}
\end{figure}

\begin{figure}
\caption{$\mu_s/M_G$ vs.\ $\xi_G\equiv 1/M_G$.  The dashed line
represents the continuum result from Monte Carlo data.}
\label{scaling}
\end{figure}

\begin{figure}
\caption{Asymptotic scaling test by using strong-coupling estimates.
The dotted line represents the exact result (\protect\ref{mass-lambda}).}
\label{asySC}
\end{figure}

\end{document}